**Routine Usage of AI-based Chest X-ray Reading Support in a Multi-site Medical Supply Center**

Authors: **Karsten Ridder[1], MD / Alexander Preuhs[2], PhD / Axel Mertins[1] / Clemens Joerger[2], PhD**

Affiliations: **1: MVZ Uhlenbrock&Partner, Dortmund, Germany**

**2: Siemens Healthineers, Erlangen, Germany**

Keywords: **Artificial Intelligence, Chest X-ray, Radiology**

Key information:

1. Research question: How can we establish an AI support for reading of chest X-rays in clinical routine and which benefits emerge for the clinicians and radiologists. Can it perform 24/7 support for practicing clinicians?
2. Findings: We installed an AI solution for Chest X-ray in a given structure (MVZ Uhlenbrock & Partner, Germany). We could demonstrate the practicability, performance, and benefits in 10 connected clinical sites.
3. Meaning: A commercially available AI solution for the evaluation of Chest X-ray images is able to help radiologists and clinical colleagues 24/7 in a complex environment. The system performs in a robust manner – supporting radiologists and clinical colleagues in their important decisions– in practises and hospitals regardless of the user and X-ray system type producing the image-data.

Introduction

As one of the worldwide most taken X-ray procedures, chest X-ray (CXR) is one of the most important and also most demanding modalities in daily medical imaging. Due to its accessibility and fast acquisition, chest radiographs are frequently used as a first diagnostic step, deciding on further patient treatment [1,2]. During normal working times there is a high workload for every radiologist with a certain risk of volatile errors, missed diagnosis, misinterpretation, and late diagnosis. While most of the research assessing AI-based clinical decision support systems is focused on quantitative evaluations in a retrospective and synthetic environment [3,4], this abstract is focused on the implementation of an AI-based chest evaluation system in a real-world clinical setup. Therefore, we deployed an AI system, capable of analysing CXR, in a large clinical setup including 10 clinical sites and qualitatively assess the benefits of such a system to the clinicians and radiologists.

Material and methods

We installed an AI-based CXR reading support in clinical routine in a German multi-site medical supply centre (MVZ Uhlenbrock and Partner). The MVZ for radiology connects 14 institutions including hospitals and private practices. 10 of these institutions produce CXR images in a clinical routine set-up 24/7. All acquired posterior anterior (PA) images from stationary X-ray systems of different vendors (Siemens Healthineers, Carestream Health, Oehm und Rehbein GmbH, Philips) are sent to a central server where they are uploaded to the cloud for evaluation. We use the AI Rad Companion Chest X-ray from Siemens Healthineers (Figure 1) [5]. The AI system identifies five pre-specified findings on CXRs (p.a. view): Pneumothorax, Pleural Effusion, Pulmonary Lesions, Consolidation, and Atelectasis. To assess the impact of the deployed AI system we retrospectively analysed the AI-detected findings, and we conducted several interviews with radiologists and connected physicians (e.g. surgeons and internal doctors) about the benefits.

Results

Within one month about 1400 frontal images, received from the 10 institutions, were analysed. Figure 2 shows the percentage of CXR images with one or more findings. Depending on the institution the images containing radiographic findings vary between 22% and 53% (Figure 3). The radiologists at the respective sites reassured that those numbers are realistic and are due to the different patient population that are treated (out-patient vs hospital). Network performance did not slow down the clinical flow of reading and the AI result matches the physician's opinion in 8 out of 10 cases. Due to missing ground-truth, a quantitative evaluation is currently not possible. Quality improvement by a kind of second read and diagnostic support during extra hours (e.g. nightshifts) were mentioned as the most important benefits during the interviews. The AI was considered especially helpful by the radiologists, if it detects difficult findings, e.g., as depicted in Figure 4, where the AI highlights a lesion behind the heart, which was confirmed later by a CT scan as shown in Figure 5. We received mixed feedback about preferred positive predictive value (PPV) and negative predictive value (NPV), with a slight tendency towards high NPV for radiologists and high PPV for clinicians. Finally, we learnt that training and communication of how to use the AI in daily routine by the users is an important success factor.

Discussion and Conclusion

To the best of our knowledge, we were the first to deploy an AI system for CXR interpretation on a large scale to a multi-centre clinical care setting. We showed that such systems can be deployed without interrupting the clinical workflow, while bringing big benefit to the physicians and the affected patients, e.g., by quality improvement through 24/7 availability and highlighting of potential abnormalities. The biggest advantage reported by the physicians was the comfort of a "second look by the AI". We observed different preferred PPVs and NPVs for clinicians and radiologists. High NPV allow expert radiologists to be more efficient in reporting, whereas high PPVs are preferred by clinicians to not oversee anything.

Disclosures

Alexander Preuhs and Clemens Joerger are employed by Siemens Healthineers

*APPENDIX*

# Figures and pictures.

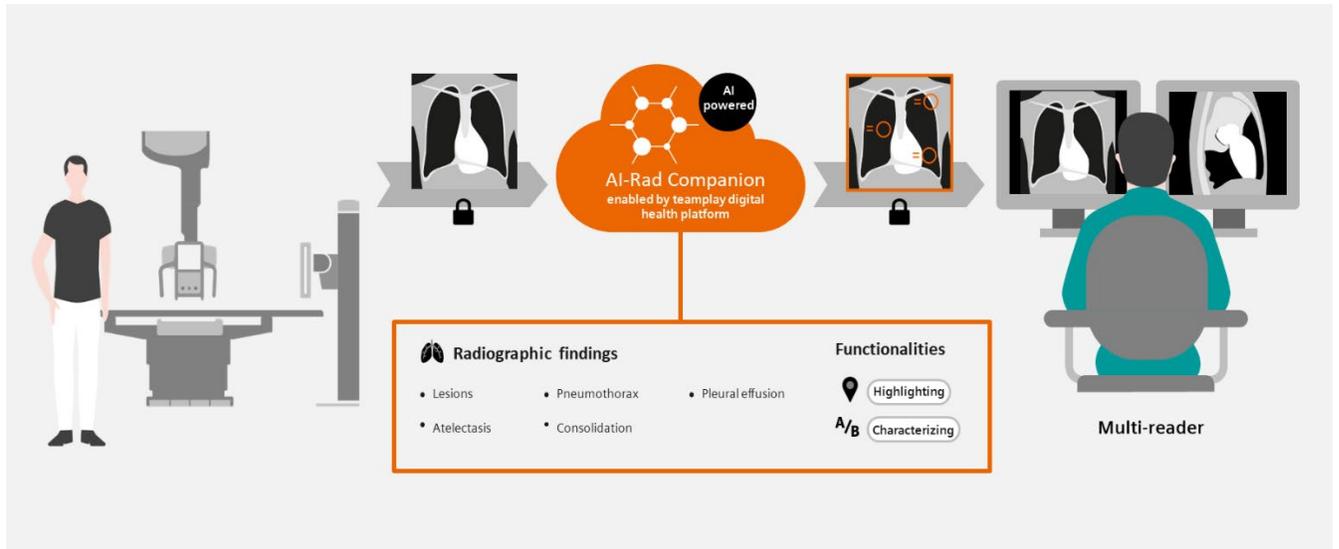

*Figure 1: Set-up of the AI Rad Companion Chest X-ray. First, a chest X-ray is acquired on a standard X-ray machine. Then, the image is securely transferred to the AI-Rad Companion which is deployed as a cloud service. There, the image is analyzed and a secondary capture of the identified findings is sent back to the reading station.*

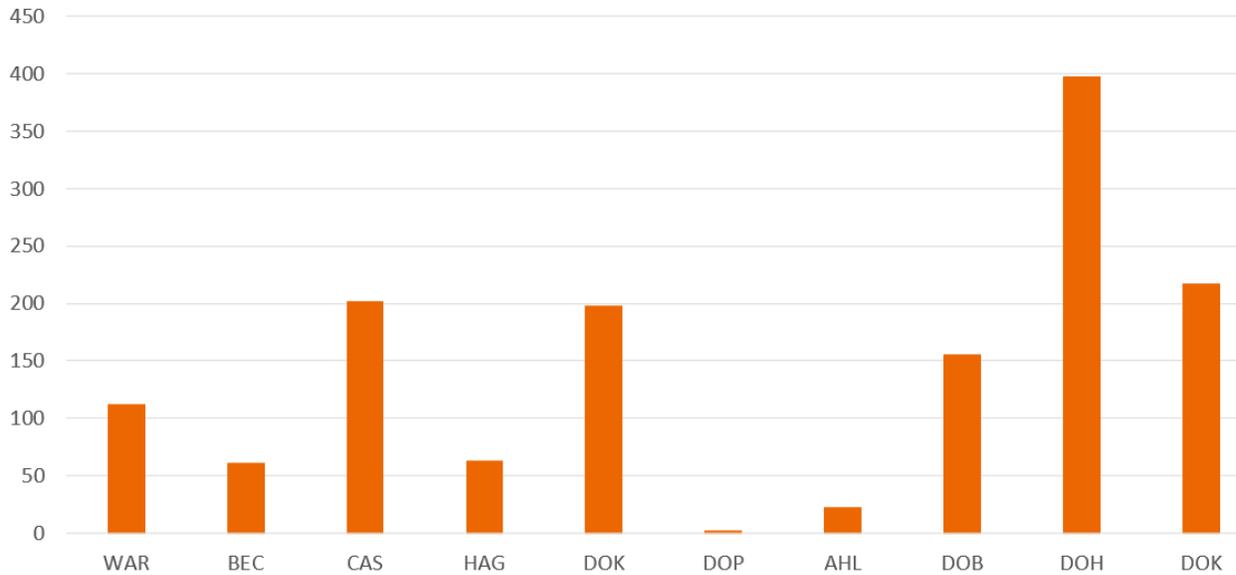

*Figure 2: Distribution of 1400 analyzed images of AI Rad Companion Chest X-ray (p.a. view) within one month across the 10 connected institutions.*

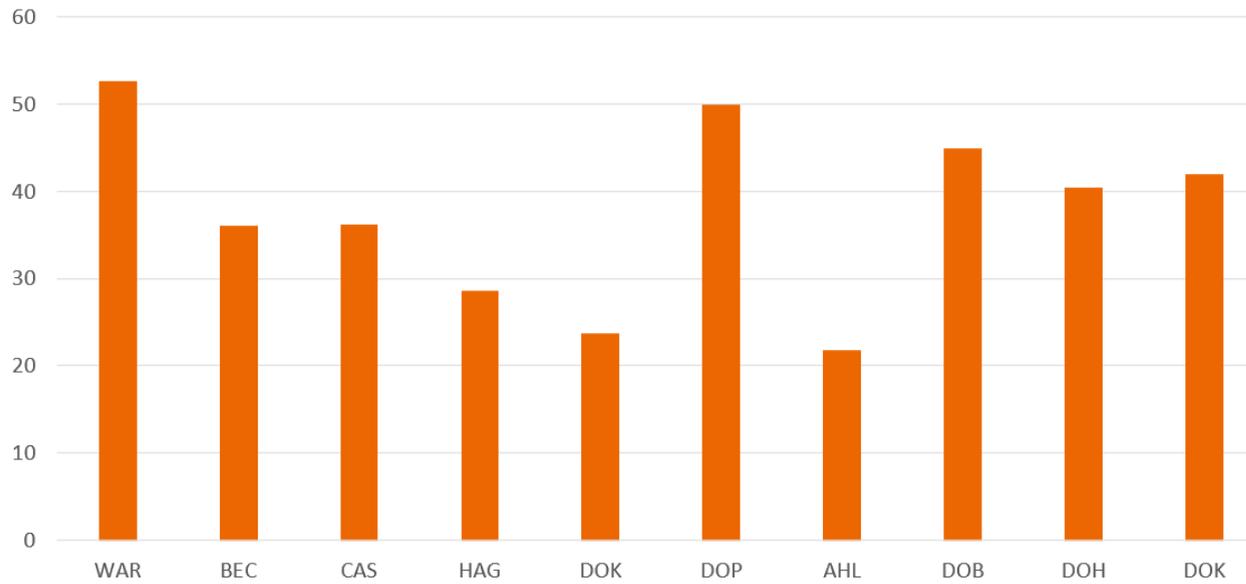

Figure 3: Percentage of AI-results of CXR (p.a. view) with at least one clinical findings across the 10 connected institutions.

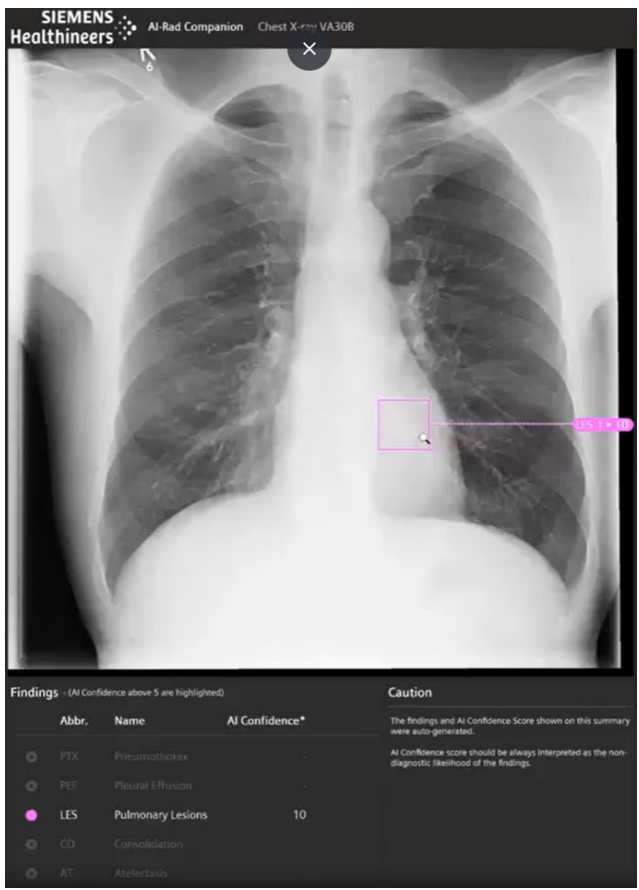

Figure 4: Exemplary secondary capture image of the AI-Rad Companion Chest X-ray. This result image indicates that the AI-algorithm found a Pulmonary lesion, whereas none of the other four findings where detected. With the AI-detected lesion being located behind the heart, this lesion could potentially be overseen in clinical routine. A CT scan of the patient confirmed the finding as displayed in Figure 5.

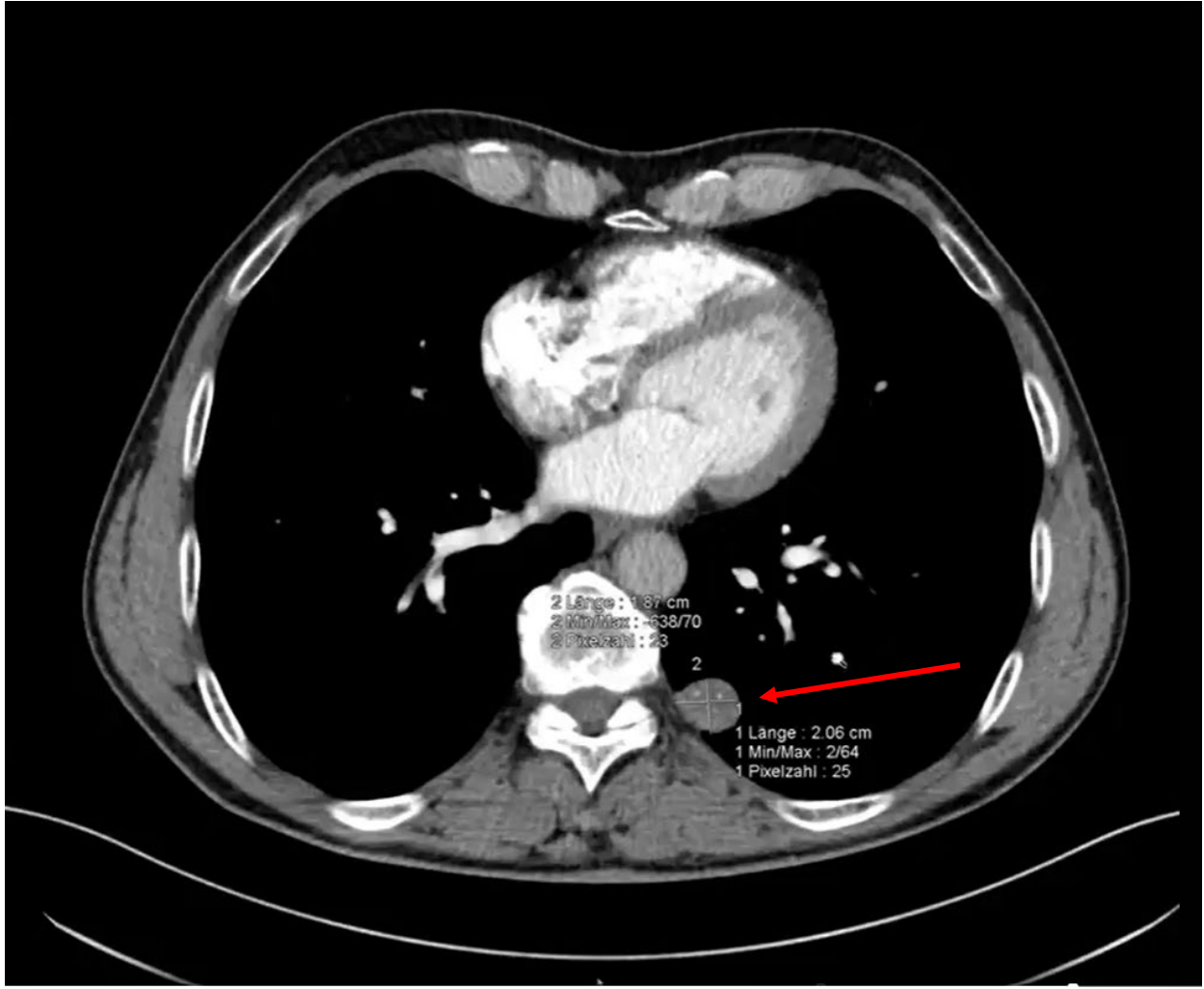

*Figure 5: Chest CT confirming pulmonary lesion (red arrow) which was identified correctly by the AI-algorithm, as displayed in Figure 4.*